\documentclass[%
 reprint,
 amsmath,amssymb,
 aps,
]{revtex4-1}
\usepackage{graphicx}
\usepackage{dcolumn}
\usepackage{bm}
\usepackage{hyperref}
\usepackage{ragged2e}
\usepackage[disable,colorinlistoftodos]{todonotes}

\usepackage{amsfonts}

\begin{document}


\title{Signatures of Topological Superconductivity and Parafermion Zero Modes in Fractional Quantum Hall Edges}

\author{Noam Schiller}
\author{Eyal Cornfeld}
\author{Erez Berg}
\author{Yuval Oreg}
 \affiliation{Department of Condensed Matter Physics, Weizmann Institute of Science, Rehovot 76100, Israel}

\date{\today}

\begin{abstract}
\justify
Parafermion zero modes are exotic emergent excitations that can be considered as $\mathbb{Z}_n$ generalizations of Majorana fermions. Present in fractional quantum Hall-superconductor hybrid systems, among others, they can serve as potential building blocks in fault-tolerant topological quantum computing. We propose a system that reveals noise and current signatures indicative of parafermion zero modes. The system is comprised of the edge excitations (``quasi-particles") of a fractional quantum Hall bulk at filling factor $\nu = 1/m$ (for odd integer $m$) incident upon the interface of a superconductor, and in the presence of back-scattering. Using perturbative calculations, we derive the current that propagates away from this structure, and its corresponding noise correlation function. Renormalization group analysis reveals a flow from a UV fixed point of perfect normal reflection towards an IR fixed point of perfect Andreev reflection. The power law dependence of the differential conductance near these fixed points is determined by $m$. We find that behavior at these two limits is physically distinguishable; whereas near perfect Andreev reflection, the system deviates from equilibrium via tunneling of Cooper pairs between the two edges, near perfect Normal reflection this is done via tunneling of a single quasi-particle to an emergent parafermion zero mode at the superconductor interface. These results are fortified by an exact solution.
\end{abstract}
\maketitle

\section{Introduction}
\label{Introduction}

Recent years have seen a bevy of interest in topological phases and any measurable quantities they may display. Chief among such phases in strongly interacting systems is the fractional quantum Hall effect (FQH)~\cite{wen_quantum_2004,jain_2007}. Such systems, which can be constructed by placing a two-dimensional electron gas in a perpendicular magnetic field, give rise to excitations (``quasi-particles") with fractional charges, determined by the filling factor $\nu=\frac{h n}{eB}$, where $h$ is the Planck constant, $n$ is the electron density, $e$ is the electron charge, and $B$ is the magnetic field. A bulk-boundary correspondence manifests in these excitations being gapless along the edge of the sample, where they propagate with a two-terminal conductance of $G=\frac{e^2}{h} \nu$.

\todo{\tiny I'd be happy to hear any other ideas you have for references - this part feels like it needs more.- This part is fine now YO}
Among the rich phenomena embedded in FQH phases is their interplay with superconductivity, which may lead to the emergence of parafermion zero modes (PZMs)~\cite{lindner_fractionalizing_2012,clarke_exotic_2013}. These exotic excitations can be considered as $\mathbb{Z}_n$ generalizations of Majorana fermions with $n>2$~\cite{fendley_parafermionic_2012,alicea_topological_2016}. Much like Majorana fermions, the non-locality of PZMs serves as protection from de-coherence effects, making PZMs potential candidates of serving as non-Abelian building blocks for topological quantum computing~\cite{stern_topological_2013}. PZMs have the additional benefit of a larger ground state degeneracy, which can not be split by electron tunneling, and as such are even more robust than Majorana fermions. Several realizations of PZMs have been suggested~\cite{alicea_topological_2016}. We propose an experimental setup revealing indicative signatures of these PZMs.

In this work, we analyze the scattering of FQH quasi-particles at filling factor $\nu = \frac{1}{m}$ (for odd integer $m$) from the interface of a superconductor, in the presence of back-scattering, via the proposed setup in Fig. \hyperref[fig:System]{\ref{fig:System}(a)}. In it, we proximity-couple two such states to a superconductor with a large gap, and allow back-scattering in a small region immediately preceding the superconductor. The upper, right-moving edge is placed at a bias voltage$~V$, producing a propagating edge current of $I_R=\frac{e^2V}{hm}$ which is incident upon the superconducting interface, at which a PZM emerges. We are interested in calculating the left-moving current which is scattered from this interface.
\todo{\tiny Yuval, you proposed re-wording this paragraph to that the SC needs SOC and that the space between the edges is a trench - I believe this is better left for the second paragraph of section II- OK YO}

For electrons ($m=1$) this scattering is well understood~\cite{haim_signatures_2015,nilsson_splitting_2008,anantram_current_1996} -- a single electron can not enter the superconductor at energies under the superconducting gap, so it may be either normally reflected, or, Andreev reflected when it enters the superconductor paired with another electron. The current can hence be calculated exactly using a scattering matrix approach~\cite{nilsson_splitting_2008}, and has been measured experimentally~\cite{lee_inducing_2017}. Quasi-particles of the fractional quantum Hall phases may undergo more complex processes. We investigate these processes by calculating the current that will be measured propagating away from the interface with the superconductor, as well as its corresponding noise correlation function. 

Analysis of this configuration gives rise to familiar ultraviolet (UV) and infrared (IR) fixed points of perfect normal and Andreev reflection, respectively~\cite{fidkowski_universal_2012,lutchyn_transport_2013}. Behavior near these fixed points is analyzed perturbatively using the real-time Keldysh technique, revealing a power law dependence of the deviation from equilibrium current: $ I + \frac{e^2 V}{h m} \propto V^{4m-1}$ near perfect Andreev reflection, and $I - \frac{e^2 V}{h m} \propto V^{\frac{1}{m}-1}$ near perfect normal reflection.

At these limits, the Fano factor $\frac{S}{2 (I \pm \frac{e^2 V}{h m})}$, where the plus (minus) refers to near-perfect Andreev (normal) reflection, and $S$ is the noise correlation function, can be interpreted as the basic charge that tunnels between edges in this process~\cite{de-picciotto_direct_1998,martin_noise_2005, chamon_tunneling_1995, kane_nonequilibrium_1994}, allowing an experimentally accessible way of probing our proposed configuration. We show that this ratio is $2e$ for the IR fixed point, and $\frac{e}{m}$ at the UV fixed point, exposing a stark physical contrast between these two limits: whereas at zero energy, the system deviates from equilibrium via Cooper pairs that the superconductor ``loses" to the exiting edge mode, at high energies these deviations arise from fractional quasi-particles tunneling from the edges onto the superconductor interface. As the superconductor itself can support no such excitations, these charges must be absorbed by the PZM. Measurement of the predicted Fano factor at high energies thus serves as an experimental signature of these modes.

Congruent with these perturbative calculations is an exact solution, similar to the one introduced in Refs.~\cite{fendley_exact_PRL,fendley_exact_PRB}, using tools of integrable systems and the thermodynamic Bethe ansatz (TBA). At finite temperature, this solution can be obtained only for perfect normal reflection. However, at zero temperature a duality exists between the two limits, enabling a valid solution for the entire energy range. We present the solution explicitly for $m=3$ only.

This manuscript is organized as follows. Section \ref{System Description} contains a description of our system. Section \ref{Calculation methodology} contains an overview of the calculation methodology used in both the perturbative and exact solutions. Sections \ref{Near_Perfect_Andreev} and \ref{Near_Perfect_Normal} describe the two limits at which perturbative calculations were done. Section \ref{Exact Solution} describes the exact solution, and the limits at which it is valid.

\section{System Description}
\label{System Description}
\begin{figure}
    \centering
    \includegraphics[width=0.5\textwidth]{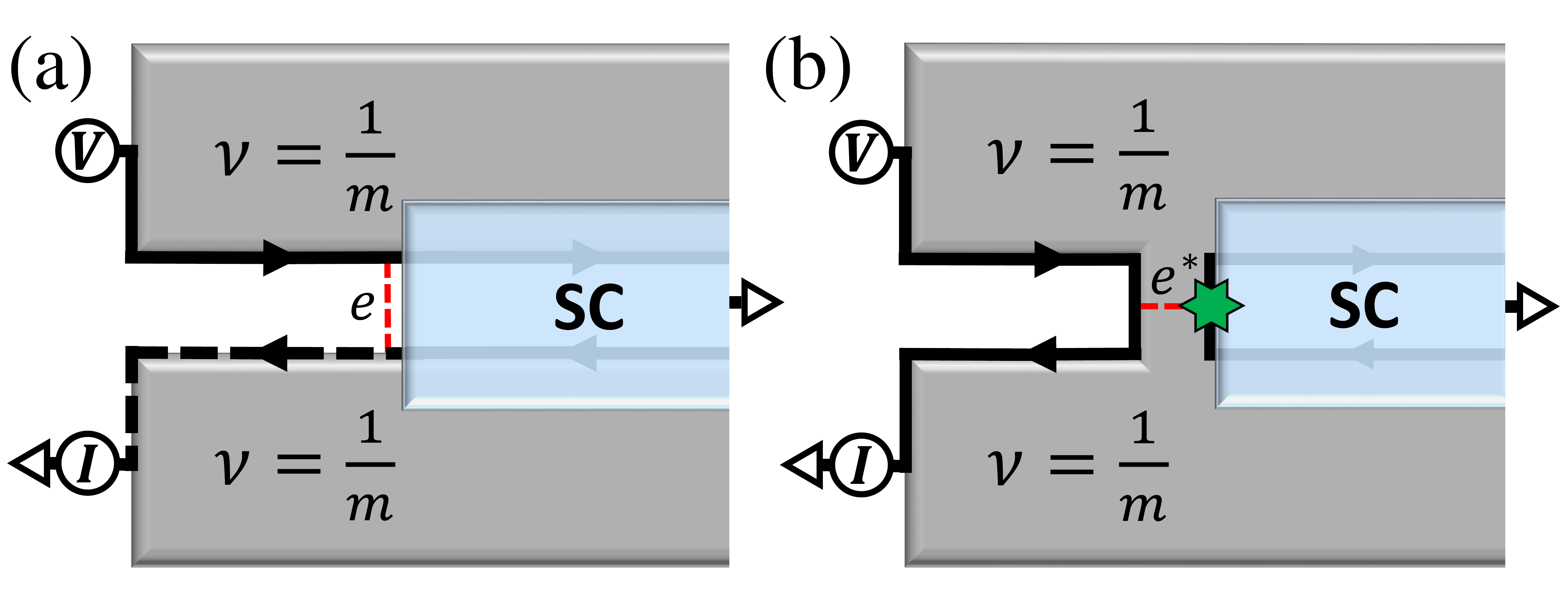}
     \caption{The system we analyze in this manuscript. Two FQH states at filling factor $\nu = \frac{1}{m}$ are proximity coupled to a half-infinite superconductor (SC). The upper, right-moving edge is placed at a bias voltage V. The lower, left-moving edge and the SC are grounded. In (a), the system is near perfect Andreev reflection - the propagating quasi-particles (full line) are reflected as quasi-holes (dashed line). The two chiral FQH edge modes are separated by vacuum. The lowest order perturbation is back-scattering of a right-moving electron to a left-moving electron. In (b), there is full back-scattering between the edges, such that they are effectively one edge of propagating quasi-particles separated from the SC by a FQH bulk. The lowest order perturbation is tunneling of an $e^*=\frac{e}{m}$ quasi-particle to a parafermion zero mode (PZM) at the SC's interface (denoted by the star).}
     \label{fig:System}
\end{figure}

We consider the system described in Fig. \hyperref[fig:System]{\ref{fig:System}(a)}. It is comprised of two separate FQH states, each at filling factor $\nu=\frac{1}{m}$, with odd integer $m$. The gapless excitations of these states are chiral edge modes carrying a charge of $e^*=\frac{e}{m}$. We assume no interactions between the two edges. Proximity coupled to both edge modes is a half-infinite superconductor. Back-scattering between the edges is allowed in a small region to the left of the superconducting interface. The upper, right-moving edge is coupled to a lead with a bias voltage of $V$, whereas both the lower, left-moving edge and the superconductor are grounded. We assume that all relevant energies are lower than the superconducting gap, such that charge can enter the superconductor only via Cooper pairs.

The precise implementation of this mechanism depends on the spin polarization of the two edge modes. If the two edge modes have opposite spins, a ferromagnet or an insulating barrier with spin-orbit coupling is required to induce the back-scattering, and the Cooper pairing of the edges may be achieved via an $s$-wave superconductor. If, however, the two edge modes have the same spins, back-scattering does not require a ferromagnet, but the pairing requires using a superconductor with strong spin-orbit coupling.

This system can be described using the Hamiltonian~\cite{wen_quantum_2004,wen_chiral_1990,giamarchi_quantum_2004} ${H}=H_0+H_B+H_{\Delta}$, where
\begin{equation}
    \label{Unperturbed_Hamilonian}
    H_0=\frac{vm}{2\pi}\int_{-\infty}^{\infty}dx[(\partial_x \varphi)^2+(\partial_x \theta)^2]
\end{equation}
is the gapless edge Hamiltonian, and 
\begin{gather}
    \label{Electron_tunneling}
    H_B=\int_{-a}^{0}dx \left[ \frac{B}{2} e^{i2m\varphi(x)} + h.c. \right],\\
    \label{SC_pairings}
    H_{\Delta}=\int_{0}^{\infty}dx \left[ \frac{\Delta}{2} e^{i2m\theta(x)} + h.c. \right],
\end{gather}
describe the back-scattering and superconducting proximity couplings, respectively. The width of the back-scattering region is $a$, whereas the width of the superconductor is infinite. Here $\theta$ and $\varphi$ are boson fields defined such that a right or left-moving electron is described by the operators $\psi_{R/L} \sim e^{im( \theta  \pm \varphi)}$. The two boson fields satisfy the commutation relation $[\theta(x),\varphi(x')]=i\frac{\pi}{m}\Theta(x'-x)$ where $\Theta$ is the Heavyside step function. Right or left-moving electron density is hence given by $\rho_{R/L}=-\frac{1}{2\pi}(\partial_x \varphi \pm \partial_x \theta)$.

We take the superconducting pairing amplitude to be infinite, $\Delta\rightarrow\infty$. As such, Eq.~\eqref{SC_pairings} dictates that the bosonic field $\theta$ is pinned to a constant value in the half-infinite domain $[0,\infty]$. This pinning reduces the integration domain in Eq.~\eqref{Unperturbed_Hamilonian} to $[-\infty,0]$, and imposes the boundary condition $\partial_t\theta(x=0,t)=0$. Additionally, we take the limit of a point-like back-scattering region, i.e. $a \rightarrow 0$.

A similar setup is shown in Fig. \hyperref[fig:System]{\ref{fig:System}(b)}. We have the same edge modes and superconductor as in Fig. \hyperref[fig:System]{\ref{fig:System}(a)}, but the two edge modes are connected such that they are separated from the superconductor by a thin area of a FQH state. As such, the relevant tunneling operator is of fractional quasi-particles to the PZM that emerges at the interface~\cite{lindner_fractionalizing_2012,clarke_exotic_2013}. 

Remarkably, while the two configurations are physically distinguishable, at zero temperature they are dual~\cite{fendley_exact_PRB}, and both converge to one of the two asymptotic fixed points of our model. At the limit $B \rightarrow 0$ (henceforth ``perfect Andreev reflection"), we identify the term $H_B$ as electron tunneling at $x=0$, giving the behavior of Fig. \hyperref[fig:System]{\ref{fig:System}(a)}. 

Conversely, at the limit $B \rightarrow \infty$ (henceforth ``perfect normal reflection"), the low energy excitations are kinks in the boson field $\varphi$ between adjacent minima of $H_B$. From the bosonic commutation relations, we find that the operator $e^{-i\left(\theta(0^+)-\theta(0^-)\right)}$ shifts $\varphi(0) \mapsto \varphi(0)+\frac{\pi}{m}$ and thus moves the field between minima. This is precisely the operator that describes tunneling of a quasi-particle from the edge modes to the PZM that emerges at the superconductor's interface~\cite{lindner_fractionalizing_2012,clarke_exotic_2013}. We thus identify perturbative tunneling of quasi-particles with a near-infinite electron tunneling amplitude. Since $\theta(0^+)$ is pinned, the interaction term $H_B$ from Eq.~\eqref{Electron_tunneling} becomes
\begin{equation}
    \label{QP_tunneling}
    H_{\tilde{B}}=\frac{\tilde{B}}{2}e^{i\theta(0^-)}+h.c.,
\end{equation}
where $\tilde{B}$ is the tunneling amplitude. This interaction term describes the quasi-particle tunneling shown via the dashed line in Fig. \hyperref[fig:System]{\ref{fig:System}(b)}. The renormalization group (RG) flows of the tunneling amplitudes $B,\tilde{B}$ are given to first order by the equations
\begin{equation}
\label{RG_flow}
        \frac{dB}{dl}=(1-2m)B, \quad
        \frac{d\tilde{B}}{dl}=(1-\frac{1}{2m})\tilde{B}.
\end{equation}

Since $m$ is an odd integer, $B$ is irrelevant and $\tilde{B}$ is relevant. As such, we can identify the perfect Andreev reflection and perfect normal reflection limits as the IR and UV limits of our system, respectively. Each of these two interactions introduces a crossover scale,
\begin{equation}
\label{Crossover_scales}
    T_B \propto B^{\frac{1}{1-2m}}, \quad
    T_{\tilde{B}} \propto \tilde{B}^{\frac{2m}{2m-1}}.
\end{equation}

As the only relevant energy scale is the crossover between the two limits, we can identify these as the same scale, and hence $B \sim \tilde{B}^{-2m}$. 

\section{Calculation methodology}
\label{Calculation methodology}

We wish to compute the left-moving current at the bottom reservoir, as well as its noise correlation. To do this, we note that the total amount of charge carriers on the right- and left-moving edges is given by the operators
\begin{equation}
    \label{Total_charges}
    N_{R/L}=\int_{-\infty}^0dx\rho_{R/L}.
\end{equation}
Charge conservation dictates that charge carriers may only enter the right-moving edge through the lead, and exit the edge at the interface. Similarly, charge carriers may only enter the left-moving edge at the interface and deplete through the reservoir. As such, the operators $N_{R/L}$ satisfy
\begin{equation}
    \label{Total_charges_change}
	e \partial_t \hat{N}_{R/L}=-i e [N_{R/L},H]=\pm j_{R/L}(-\infty) \mp \hat{I}_N - \frac{1}{2} \hat{I}_A,
\end{equation}
where $j_{R/L}(x) = \pm e v \rho_{R/L}(x)$ are the local right/left-moving current operators, $I_N$ is the current that tunnels from the right-moving edge to the left-moving edge, and $I_A$ is the Andreev current that enters the superconductor. In a steady state, $\partial_t N_{R/L}= 0$ \footnote{This is valid in the DC limit. Indeed, when calculating noise correlation functions, we will always take $\omega \rightarrow 0$}. The bias voltage imposes $j_{R}(-\infty) = \frac{e^2V}{hm}$; this allows us to write the current along the left edge as
\begin{equation}
    \label{Left_moving_current}
    \hat{I}(t) = j_{L}(-\infty) = 2 \hat{I}_N - \frac{e^2V}{hm}.
\end{equation}
Charge conservation along the edge dictates that this operator will describe the current depleting at the reservoir as well. We write this operator in terms of the bosonic fields using Eq.~\eqref{Total_charges_change}, and its expectation value is then calculated perturbatively for the two limits described in Fig. \ref{fig:System} using the real-time Keldysh formalism~\cite{kamenev_many-body_2004}. 

The unperturbed Hamiltonian described in Eq.~\eqref{Unperturbed_Hamilonian} is used to derive the Keldysh-rotated Green's functions, as elaborated in \hyperref[app:Keldysh]{Appendix A}. For each of the limits, the lowest order tunneling operator is then treated perturbatively; near perfect Andreev reflection (Fig. \hyperref[fig:System]{\ref{fig:System}~(a)}), this will be the electron tunneling operator (Eq.~\eqref{Electron_tunneling}), whereas near perfect normal reflection (Fig. \hyperref[fig:System]{\ref{fig:System}(b)}), this will be the quasi-particles tunneling operator (Eq.~\eqref{QP_tunneling}). The bias voltage on the right-moving edge is introduced via a gauge transformation $\psi_R\rightarrow\psi_R e^{i\omega_0t}$, where $\omega_0=eV$ (taking natural units $\hbar=1$); the way $\psi_L$ transforms will be dictated by the limit at which the calculation is performed (see Eqs.~\eqref{Andreev_Gauge_Transformation},~\eqref{Normal_Gauge_Transformation}).

The exact solution is obtained by mapping our system to that described by Fendley, Ludwig and Saleur~\cite{fendley_exact_PRL,fendley_exact_PRB}, and following their methodology with the appropriate modifications, as elaborated in \hyperref[app:Exact]{Appendix B}.

In both methodologies, the observables that are of interest to us are the left-moving current $\hat{I}(t)$, its corresponding differential conductance $G=\frac{dI}{dV}$, and its noise correlation function
\begin{equation}
    \label{Noise_Definition}
    S(\omega)=\int_{-\infty}^{\infty} dt 
    e^{i\omega t} \langle \hat{I}(t)\hat{I}(0)+\hat{I}(0)\hat{I}(t) \rangle.
\end{equation}

\section{Near Perfect Andreev Reflection}
\label{Near_Perfect_Andreev}
Near perfect Andreev reflection, the interaction term which we treat perturbatively is the electron tunneling operator (Eq.~\eqref{Electron_tunneling}). As such it will be convenient to describe the action $S=S_0+S_{B}$ in terms of the bosonic field~$\varphi$ (Eq.~\eqref{Unperturbed_Actions_Phi}). At zero order, all incident right-moving particles are reflected as left-moving holes. Therefore, the left-moving edge equilibrates at a voltage of $-V$. We can hence conveniently treat the voltages of both edges via the gauge transformation
\begin{equation}
    \label{Andreev_Gauge_Transformation}
    \psi'_{R/L}=\psi_{R/L}e^{\pm i\omega_0 t},\quad\omega_0=eV,
\end{equation}
under which $B$ and $\varphi$ transform as
\begin{equation}
    \label{Andreev_Gauge_Transformation_B_Phi}
    B'=Be^{-2 i\omega_0 t},\quad \varphi'=\varphi+\frac{\omega_0 t}{m}.
\end{equation}
After calculating the appropriate commutators, Eq.~\eqref{Left_moving_current} gives the left-moving current operator
\begin{equation}
    \label{Left_Moving_Current_Operator_Andreev}
    \hat{I}(t)= -\frac{e^2V}{hm}+
    \left[ \frac{eB}{i}e^{-2i \omega_0 t}e^{2im\varphi'(t)}+\text{h.c.} \right],
\end{equation}

The zero-order expectation value of the current is trivially $-\frac{e^2V}{hm}$, since in the absence of electron tunneling all incident particles are perfectly Andreev reflected. Defining the deviation from this point as $\delta\hat{I}(t) \equiv \hat{I}(t) +\frac{e^2V}{hm}$, we proceed to calculate the expectation value of $\delta \hat{I}$ perturbatively in the interaction term $H_B$. This can be done both in zero and finite temperature, using the correlation functions calculated  \hyperref[app:Keldysh]{Appendix A}.

Charge conservation dictates that only even powers of $B$ contribute. A direct calculation shows that to order $O(B^2)$, the zero-temperature current is
\begin{equation}
    \label{Current_Andreev_Zero_T}
    \langle \delta \hat{I}(t) \rangle= \frac{e \pi B^2}
    {\Lambda^{4m}\Gamma(4m)} \left( 2eV \right)^{4m-1},
\end{equation}
where $\Lambda$ is a UV cutoff. This is consistent with previous results for back-scattering in Luttinger liquids~\cite{kane_transport_1992,kane_transmission_1992,chamon_tunneling_1995,lutchyn_transport_2013,fidkowski_universal_2012}, with the typical Luttinger parameter $K$ replaced here by $2m$. Higher orders depend on the voltage via 
$O(B^{2k}) \propto \frac{1}{\Lambda^{4km}}\left( 2eV \right)^{4km-(2k-1)}$, but we eschew explicit calculation of the coefficients here. For $m=1$, these results are further consistent with the scattering matrix solution to the electron case~\cite{nilsson_splitting_2008}

\begin{equation}
    \label{Electron_Solution}
    \langle \hat{I}(t) \rangle = \frac{e^2V}{h} 
    \left[ 1 - 2\frac{T_B}{eV} \tan^{-1} \left( \frac{eV}{T_B} \right) \right].
\end{equation}

Using the same methods, we calculate the DC noise contribution
\begin{equation}
    \label{Noise_Andreev_Zero_T}
    S(\omega \rightarrow 0)= \frac{4 e^2 \pi B^2} {\Lambda^{4m}\Gamma(4m)} 
    \left( 2eV \right)^{4m-1} = 4e \langle \delta \hat{I}(t) \rangle,
\end{equation}
giving us a Fano factor of
\begin{equation}
    \label{Fano_Andreev_Zero_T}
    \frac{S}{2 \langle \delta \hat{I}(t) \rangle}=2e.
\end{equation}
This is consistent with our understanding of the system. Since incident electrons can not enter the superconductor, they must be reflected as either left-moving electrons or left-moving holes, with the difference between these two outcomes being a Cooper pair. Interpreting the Fano factor as the basic unit of charge being transferred by the interaction term indeed corresponds to such pairs.

Repeating the calculations above for finite temperature $T$ gives a current of
\begin{align}
     \label{Current_Andreev_Finite_T}
    \langle \delta \hat{I}(t) \rangle = &\frac{e B^2}
    {\Lambda} \left( \frac{2\pi T}{\Lambda}\right)^{4m-1}
    \sinh \left( \frac{eV}{T}\right) \times \\
    & \mathcal{B} \left( 2m+i\frac{eV}{\pi T}, 2m-i\frac{eV}{\pi T} \right) \nonumber,
\end{align}
where $\mathcal{B}(x,y)$ is the beta function. This coincides with Eq.~\eqref{Current_Andreev_Zero_T} for $\frac{eV}{T} \gg 1$. At the limit $\frac{eV}{T} \ll 1$, we obtain
\begin{equation}
     \label{Current_Andreev_Infinite_T}
    \langle \delta \hat{I}(t) \rangle \approx \frac{2 \pi e^2 V B^2} {\Lambda^2}
    \mathcal{B}(2m,2m) \left( \frac{2\pi T}{\Lambda}\right)^{4m-2}.
\end{equation}
The finite-temperature DC noise contribution is
\begin{equation}
    \label{Noise_Andreev_Finite_T}
    S(\omega \rightarrow 0)= 4e \langle \delta \hat{I}(t) \rangle \coth \left( \frac{eV}{T}\right).
\end{equation}

\section{Near Perfect Normal Reflection}
\label{Near_Perfect_Normal}

The perturbative calculations in this limit are remarkably similar to the previous limit. The interaction term which we treat perturbatively is now the quasi-particle tunneling operator (Eq.~\eqref{QP_tunneling}) instead of the electron tunneling operator (Eq.~\eqref{Electron_tunneling}). As such it will be convenient to describe the action in terms of the bosonic field~$\theta$~(Eq.~\eqref{Unperturbed_Actions_Theta}). Additionally, we change our gauge transformation to
\begin{equation}
    \label{Normal_Gauge_Transformation}
    \psi'_{R/L}=\psi_{R/L}e^{i\omega_0 t},
\end{equation}
since in this limit, at zero order, all incident right-moving particles are reflected as left-moving particles, and the left-moving edge equilibrates at a voltage of $V$. Given these changes, it can be shown that the left-moving current operator in this limit is identical Eq.~\eqref{Left_Moving_Current_Operator_Andreev}, up to a minus sign on the current operator, and the transformations
\begin{equation}
    \label{transformations}
    2m \rightarrow \frac{1}{2m}, \: \theta \rightarrow 2m \varphi,
    \: e\rightarrow\frac{e}{2m}, \: B\rightarrow\tilde{B}.
\end{equation}

As such, all calculations obtained previously leading to Eq.~\eqref{Current_Andreev_Zero_T} can be modified to this limit via these transformations. The zero-order left-moving current is now $\frac{e^2V}{hm}$. Defining the differential current as $\delta \hat{I}(t) \equiv \hat{I}(t)-\frac{e^2V}{hm}$, we now obtain the zero-temperature current
\begin{equation}
    \label{Left_Moving_Current_Normal_Value_Zero_T}
    \langle \delta \hat{I}(t) \rangle= - \frac{e \pi \tilde{B}^2}
    {2m \Lambda^{\frac{1}{m}}\Gamma(\frac{1}{m})} \left( \frac{eV}{m}\right)^{\frac{1}{m}-1}.
\end{equation}

This is again consistent with previous results for Luttinger liquids near perfect normal reflection~\cite{lutchyn_transport_2013,fidkowski_universal_2012}. For $m=1$, this contribution is independent of $V$, and hence will not contribute to the differential conductance. Higher orders depend on the voltage via 
$O(\tilde{B}^{2k}) \propto \frac{1}{\Lambda ^{\frac{k}{m}}}\left( \frac{eV}{m} \right)^{\frac{k}{m}-(2k-1)}$. Notice that at certain values of $m,k$ contributions may cancel out completely -- for example, comparison to Eq.~\eqref{Electron_Solution} for $m=1$ shows that contributions for odd values $k>1$ must vanish. We again forgo explicit calculation of these coefficients.

The DC noise contribution in this limit is
\begin{equation}
    \label{NoiseNormal_Zero_T}
    S(\omega \rightarrow 0)= \frac{e^2 \pi \tilde{B}^2} {m^2 \Lambda^{\frac{1}{m}}\Gamma(\frac{1}{m})} 
    \left( \frac{eV}{m}\right)^{\frac{1}{m}-1},
\end{equation}
giving a Fano factor of
\begin{equation}
    \label{Fano_Normal_Zero_T}
    \frac{S}{2\langle \delta \hat{I}(t) \rangle} = -\frac{e}{m}.
\end{equation}
In contrast to the previous limit, the basic unit of charge being transferred by the interaction term is $\frac{e}{m}$, i.e., the particles that tunnel in our process are single quasi-particles, rather than Cooper pairs. Since the SC bulk can not support fractional quasi-particles, there must be a state at the interface that can absorb fractional charges, i.e., a PZM. We note that we obtained the effect of the PZM directly from the boundary conditions on the bosonic fields, and did not have to assume its presence.

Repeating the calculations above for finite temperature gives a current of
\begin{align}
     \label{Current_Andreev_Finite_T}
    \langle \delta \hat{I}(t) \rangle = &-\frac{e \tilde{B}^2} {2m \Lambda}
    \left( \frac{2\pi T}{\Lambda}\right)^{\frac{1}{m}-1}
    \sinh \left( \frac{eV}{2mT}\right) \times \\
    & \mathcal{B} \left( \frac{1}{2m}+i\frac{eV}{2 \pi m T}, \frac{1}{2m}-i\frac{eV}{2 \pi m T} \right) \nonumber.
\end{align}
This coincides with Eq.~\eqref{Left_Moving_Current_Normal_Value_Zero_T} at the limit $\frac{eV}{T} \gg 1$. At the opposite limit, $\frac{eV}{T} \ll 1$, we obtain
\begin{equation}
    \label{Current_Normal_Infinite_T}
    \langle \delta \hat{I}(t) \rangle \approx -\frac{\pi e^2 V \tilde{B}^2} {2 m^2 \Lambda^2}
    \mathcal{B} \left( \frac{1}{2m},\frac{1}{2m} \right)
    \left( \frac{2\pi T}{\Lambda}\right)^{\frac{1}{m}-2}.
\end{equation}
The finite-temperature DC noise contribution is
\begin{equation}
    \label{Noise_Normal_Finite_T}
    S(\omega \rightarrow 0)= - \frac{2e}{m} \langle \delta \hat{I}(t) \rangle 
    \coth \left( \frac{eV}{2mT}\right).
\end{equation}

\section{Exact Solution}
\label{Exact Solution}
In the previous sections, we presented perturbative results at the IR and UV limits. At zero temperature, we can obtain the full crossover throughout the entire energy range of the system using an exact solution. This solution was introduced in a series of papers by Fendley, Saleur, Warner and Ludwig in the early to mid-1990s~\cite{fendley_exact_PRL,fendley_exact_PRB,fendley_exact_shot_noise_1995,fendley_exact_sine_gordon,fendley_massless_1993,fendley_massless_flows_1993-1} for a quasi-particle back-scattering between two FQH edges via a point contact. 

\begin{figure}[t]
    \centering
    \includegraphics[width=0.4\textwidth]{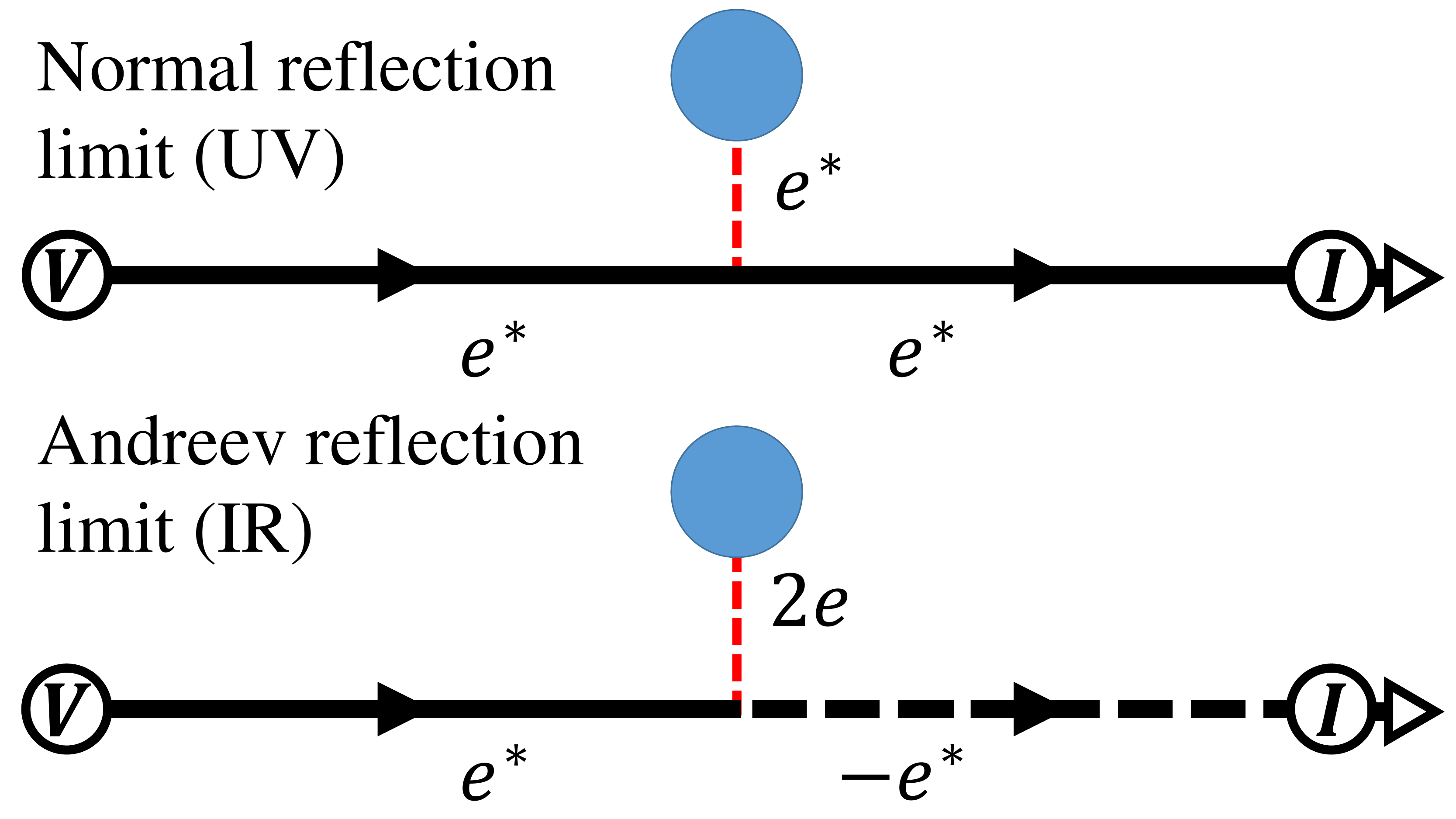}
    \caption{Schematic illustration of the mapping of the system we study to a single impurity problem. We unfurl our edge modes, such that the modes to the left (right) of the impurity are the upper (lower) edge. The superconductor interface and back-scattering are collectively treated as a single impurity. At the normal reflection limit (upper picture), both edge modes are quasi-particles. The perturbation is the impurity ``absorbing" a quasi-particle. At the Andreev reflection limit (lower picture), quasi-particles are reflected as quasi-holes. The perturbation is the impurity ``releasing" a Cooper pair in lieu of such reflection. At finite temperature, the exact solution is only valid for the upper construction.}
    \label{Chiral_Mode_Impurity}
\end{figure}

Each of the two limits described in Fig. \ref{fig:System} can be mapped onto a single, infinite chiral edge mode encountering a point-like impurity, described schematically in Fig. \ref{Chiral_Mode_Impurity}. Such a mapping was instrumental in deriving an exact solution in Ref.~\cite{fendley_exact_PRL}. It results in a Hamiltonian $H = H_0[\phi] + H_{\text{int}}[\phi]$, where $H_0[\phi]$ is the unperturbed Hamiltonian of a single chiral mode, and the interaction term is
\begin{equation}
    \label{Fendley_interaction_term}
    H_{\text{int}}[\phi] \propto \cos\left(\frac{\beta}{2} \phi(x=0)\right).
\end{equation}
The scaling dimension of this operator is $\frac{\beta^2}{8 \pi}$~\cite{fendley_exact_sine_gordon}. In~\cite{fendley_exact_PRL,fendley_exact_PRB}, this gives $\frac{\beta^2}{8 \pi} = \nu$. To complete the mapping from our configuration (with the interaction term Eq.~\eqref{QP_tunneling}) to this one, we must rescale the boson field. This results in a scaling dimension of $\frac{\beta^2}{8 \pi} = \frac{1}{2m}$ for the perfect normal reflection limit, and $\frac{\beta^2}{8 \pi} = 2m$ for the perfect Andreev reflection limit.

From here, we follow the steps in Ref.~\cite{fendley_exact_PRL,fendley_exact_PRB}, with $m=3$. As elaborated therein, at finite temperature, the resulting TBA equations are solvable for $\frac{8 \pi}{\beta^2} \in \mathbb{N}$. We hence only solve for the perfect normal reflection limit, for which $\frac{\beta^2}{8 \pi} = \frac{1}{6}$. 
\begin{figure}
    \centering
    \includegraphics[width=0.48\textwidth]{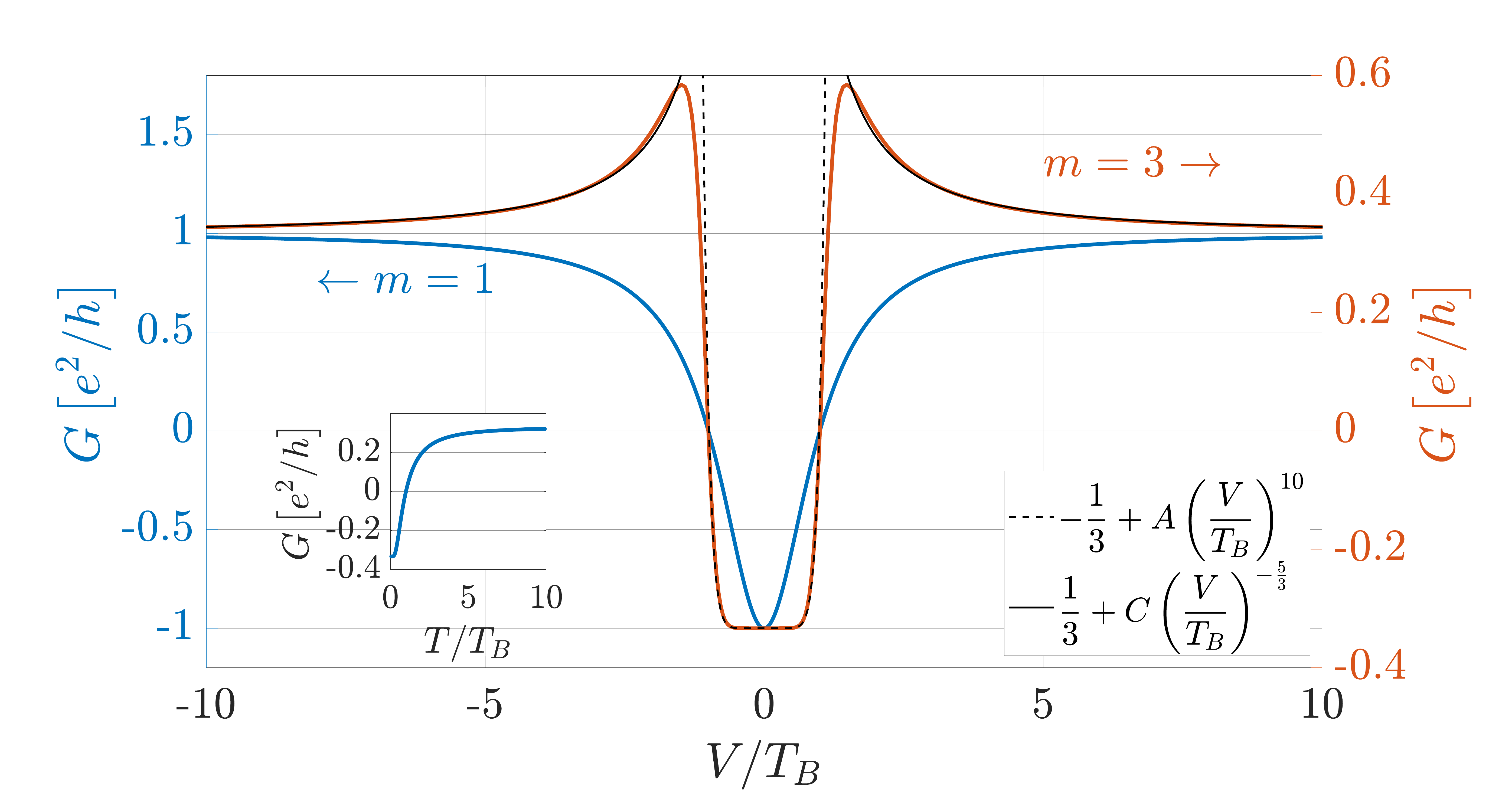}
     \caption{Differential conductance, $ G = \frac{dI}{dV} $, as obtained via solution of the TBA equations. The main plot shows the entire voltage range for zero temperature, both for the electron case ($m=1$, Eq.~\eqref{Electron_Solution}) and the $m=3$ case. The IR and UV limits agree with our perturbative calculations of $\delta G \propto \left( V/T_B \right)^{4m-2}$ and  $\delta G \propto \left( V/T_B \right)^{\frac{1}{m}-2}$ (for $m \neq 1$), respectively. The inset shows the temperature dependence for the near perfect normal reflection limit, described in Fig. \hyperref[fig:System]{\ref{fig:System}(b)}, at a voltage of $V=0$ and for $m=3$.}
    \label{dIdV_plots}
\end{figure}

The result is an exact solution for the system in Fig.~\hyperref[fig:System]{\ref{fig:System}(b)}. At zero temperature, the duality between this and Fig.~\hyperref[fig:System]{\ref{fig:System}(a)} leads to the exact solution being valid for both systems over the entire voltage range. Some details of the exact solution can be found in Appendix~\ref{app:Exact}. The differential conductance at zero temperature as a function of voltage is given in Fig. \ref{dIdV_plots}, both for the electron case ($m=1$) and the $m=3$ case. In the electron case, differentiating the current in Eq.~\eqref{Electron_Solution} gives~\cite{nilsson_splitting_2008} $G=\frac{e^2V}{h} \left[ 1-\frac{2}{1+(\frac{V}{T_B})^2} \right]$. As the scale $T_B$ is defined up to a constant, we normalize the plots such that $G=0$ at $V=T_B$. These results are consistent with our first-order perturbations: we obtain conductances of $\pm\frac{e^2}{hm}$ at the IR and UV limits, respectively. Furthermore, the first-order correction to the conductance is $\delta G \propto \left(  V/T_B \right)^{4m-2}$ in the IR limit and $\delta G \propto \left(  V/T_B \right)^{\frac{1}{m}-2}$ (for $m \neq 1$) in the UV limit -- this is shown in Fig. \ref{dIdV_plots} via the dashed and solid black curves, respectively. The constants $A$ and $C$ are obtained by fitting to the exact solution.

At finite temperature, the duality between the two limits breaks down, as thermal particle-hole excitations create quasi-particle and quasi-hole pairs in a FQH bulk vs. electron and hole pairs in the vacuum. As such, the exact solution is only valid for the normal reflection limit depicted in Fig. \hyperref[fig:System]{\ref{fig:System}(b)}. These results will converge with our system when it approaches the perfect normal reflection limit, i.e. at infinite voltage. The inset of Fig. \ref{dIdV_plots} shows the differential conductance obtained by the exact solution for system Fig. \hyperref[fig:System]{\ref{fig:System}(b)} at a voltage of $V=0$ as a function of temperature. The plot is again normalized such that $G=0$ at $T=T_B$ (notice this will give a different $T_B$ than the other plot).

\section{Discussion}

This work discusses the scattering of fractional quasi-particles from a superconductor in our proposed system (Fig. \ref{fig:System}). We consider the regime where the coupling between the superconductor and the FQH edges is sufficiently strong such that parafermion zero modes (PZMs) form at its ends. Under these conditions, at asymptotically low voltages, the system is controlled by an infrared fixed point characterized by perfect Andreev reflection, resulting in a conductance $G=dI/dV=-\frac{e^2}{h m} + \delta G$, where $V$ and $I$ are as defined in Fig.~\ref{fig:System}. $\delta G \propto \left( E/T_B \right)^{4m-2}$, with $E$ representing the dominant energy scale among $eV$ and $T$, and $T_B$ being the crossover scale of the system. As the Fano factor is $2e$, we identify the most relevant excitation of this system as tunneling of an electron to the lower edge rather than a hole -- i.e., the current depleting through the superconductor grounding loses one Cooper pair.

Conversely, at high bias voltages, the system is near perfect normal reflection. This changes the aforementioned quantities to $G=\frac{e^2}{hm}+\delta G$,  $\delta G \propto \left(  E/T_B \right)^{\frac{1}{m}-2}$ (for $m \neq 1$). The Fano factor is $\frac{e}{m}$, such that the most relevant excitation of this system is tunneling of fractional quasi-particles from the edge modes, through the bulk, and into the PZM at the superconductor's interface. 

Interestingly, the conductance can be derived exactly even away from the extreme limits of $E$ much smaller or much larger than $T_B$ using the techniques of  Refs.~\cite{fendley_exact_PRL,fendley_exact_PRB}. The exact solution describes the crossover between two limits. The features found here, including the strongly non-Lorentzian shape of the resonance in $G(V)$ (Fig. \ref{dIdV_plots}) and the crossover of the Fano factor from $2e$ at small $V$ to $\frac{e}{m}$ at large $V$, can be used to identify PZMs in experiments.

Our setup can be further generalized in a number of ways. Allowing interactions between the quasi-particles that propagate along the edges could give rise to effective Luttinger parameters; these will change the power law dependence of the current, but should not change the Fano factors at the fixed points, as they do not materially change the basic charge that is transferred. Additional deviations from these results may be obtained in systems that support more relevant tunneling processes; for example, from Eq.~\eqref{RG_flow} it follows that for $m>2$, tunneling of 2 quasi-particles through the bulk is still relevant. These will be of a lower order of magnitude than single quasi-particle tunneling, and as such will not affect the system close enough to the fixed points, but they will manifest in deviations from the exact solution. Further tunneling may be added between two edges of the superconductor, by making it of finite length or by accounting for its width. 

\section{Acknlowedgements}
This work was partially supported by the European Unions Horizon 2020 research and innovation programme (grant agreement LEGOTOP No 788715), the
DFG (CRC/Transregio 183, EI 519/7- 1), the Israel
Science Foundation (ISF), ISF and MAFAT Quantum Science and Technology grant,  and by an NSF/DMR-BSF 2018643 grant. 

\bibliography{QP_scattering_from_SC.bib}

\appendix

\section{Keldysh formalism}
\label{app:Keldysh}
\renewcommand{\theequation}{A\arabic{equation}}
\setcounter{equation}{0}

The unperturbed Hamiltonian in Eq.~\eqref{Unperturbed_Hamilonian}, after taking $\Delta \rightarrow \infty$, is given by

\begin{equation}
     H_0=\frac{vm}{2\pi}\int_{-\infty}^{0}dx[(\partial_x \varphi)^2+(\partial_x \theta)^2].
\end{equation}

Using the commutation relations between the bosonic fields, we identify two pairs of canonically conjugate variables:

\begin{equation}
\begin{split}
    \left[ \varphi(x),\frac{m}{\pi} \partial_{x'}\theta(x') \right] =i\delta(x-x') \\
    \left[ \theta(x),\frac{m}{\pi} \partial_{x'}\varphi(x') \right] =i\delta(x-x').
\end{split}
\end{equation}

Using the appropriate Legendre transformations, we can hence describe our unperturbed action entirely in terms of one of the bosonic fields~\cite{chamon_tunneling_1995}

\begin{equation}
    \label{Unperturbed_Actions_Phi}
    S_0[\varphi]=\frac{1}{2}\int_{\mathcal{C}}dt\int\limits _{-\infty}^{\infty}dx \frac{vm}{2\pi}\left[\left(\frac{1}{v}\partial_{t}\varphi\right)^{2}-\left(\partial_{x}\varphi\right)^{2}\right],
\end{equation}
\begin{equation}
    \label{Unperturbed_Actions_Theta}
    S_0[\theta]=\frac{1}{2}\int_{\mathcal{C}}dt\int\limits _{-\infty}^{\infty}dx \frac{vm}{2\pi}\left[\left(\frac{1}{v}\partial_{t}\theta\right)^{2}-\left(\partial_{x}\theta\right)^{2}\right],\\
\end{equation}
where the contour $\mathcal{C}$ is the Keldysh contour, and the factor of $\frac{1}{2}$ on the unperturbed action $S_0$ is obtained to avoid double counting when extending from the half-infinite to the infinite integration domain to $[-\infty,\infty]$ via an even continuation. This is dictated by the the pinning of $\theta(x\geq 0,t)$, which leads to the boundary conditions $\partial_t \theta(x=0,t)=\partial_x \varphi(x=0,t)=0$ and $\partial_x \theta(x=0,t)=0$.

We will continue the derivation of the Keldysh Green's functions for $\varphi$ as perform in Refs.~\cite{kamenev_2011,kamenev_many-body_2004}; calculations for $\theta$ will be identical. Explicitly separating the Keldysh contour to forward and backward propagating branches and performing a Keldysh rotation, the action is now

\begin{align}
    \label{Unperturbed_Action_Backward_Forward}
    S_0[\varphi] = & \frac{1}{2}
    \int_{-\infty}^{\infty}dtdt'\int\limits _{-\infty}^{\infty}dxdx' \times \nonumber \\
    & \vec{\varphi}^T(x,t)\hat{D}^{-1}(x,x',t,t')\vec{\varphi}(x',t'),   
\end{align}
where $\vec{\varphi}(x,t)= \left( \begin{matrix} \varphi_{cl}(x,t) \\ \varphi_{q}(x,t) \end {matrix} \right)$, using the convention $\varphi_{cl/q}=\frac{1}{2}(\varphi_+ \pm \varphi_-)$. The matrix  $\hat{D}^{-1}(x,x',t,t')$ is given via

\begin{equation}
    \label{Keldysh_matrix_real_space}
        \hat{D}^{-1}(x,x',t,t') = \delta(x-x')\delta(t-t') \left(
        \begin{matrix}
            0 & \hat{D}^{-1}_A \\
            \hat{D}^{-1}_R & \hat{D}^{-1}_K \\
        \end{matrix}
        \right).
\end{equation}
The retarted and advanced components are

\begin{equation}
    \label{Keldysh_matrix_real_components}
    \hat{D}^{-1}_{R/A}=\frac{vm}{\pi} \left[ \left( \frac{i}{v} \partial_t \pm i\varepsilon \right)^2-(i\partial_x)^2 \right].
\end{equation}
Moving to momentum space, these quantities are

\begin{align}
    \label{Keldysh_action_momentum_space}
        S_0[\varphi] = & \frac{1}{2}
        \int d\omega d\omega' \int dk dk' \delta(\omega+\omega') \delta(k+k')\times \nonumber \\
        & \vec{\varphi}^T(k,\omega)\hat{D}^{-1}(\omega,k)  \vec{\varphi}(k',\omega'),    
\end{align}
\begin{equation}
    \label{Keldysh_matrix_momentum_components}
        \hat{D}^{-1}_{R/A}(\omega,k)=\frac{1}{(2\pi)^2}\frac{vm}{\pi} 
        \left[ \left( \frac{\omega}{v} \pm i\varepsilon \right)^2-k^2 \right].
\end{equation}
Now defining $\Phi(t)=\varphi(x=0,t)$, we find the two-point correlation function via

\begin{equation}
    \label{zero_d_correlation_function}
    -i\langle \Phi_\alpha(\omega) \Phi_\beta(\omega')\rangle = \delta(\omega + \omega') \hat{D}^{\alpha \beta}
    \quad \alpha,\beta \in cl,q,
\end{equation}
giving us the retarded and advanced Green's functions

\begin{align}
    \label{zero_d_Keldysh_Green_Functions}
    \hat{D}^{R/A}(\omega) & = \int \frac{dkdk'}{(2\pi)^2} D^{R/A}(\omega,k)\delta(k+k') \nonumber \\
    & = \int \frac{dk}{vm} \frac{\pi}{\left( \frac{\omega}{v} \pm i\varepsilon \right)^2-k^2} \\
    & = \mp \frac{\pi i}{m} \frac{\pi}{\omega \pm i\varepsilon} \nonumber ,
\end{align}
where the last equality is obtained via contour integration in the complex plane. From this the Keldysh Green's function is obtainable using the relation

\begin{equation}
    \label{zero_d_Keldysh_Component}
    \hat{D}^K(\omega)= \coth \left( \frac{\omega}{2T} \right) \left( \hat{D}^R(\omega)- \hat{D}^A(\omega) \right).
\end{equation}

All relevant correlation functions are directly obtained from these quantities:

\begin{equation}
    \label{Correlation_Functions}
\begin{matrix}
    \langle \left( \Phi(t) - \Phi(t') \right)^2 \rangle = \\ 
    \begin{cases}
        \frac{2}{m} \log \left( \Lambda(\delta+i|t-t'|) \right] & T=0 \\
        \frac{2}{m} \log \left( \frac{i \Lambda}{\pi T} \sinh \left( \pi T (-i\delta+|t-t'|) \right) \right) & T>0,
    \end{cases}
\end{matrix}
\end{equation}
where $\Lambda$ and $\delta$ are UV and IR cutoffs, respectively. Since the times $t,t'$ are on the Keldysh contour, the time difference $|t-t'|$ is given by

\begin{equation}
    \label{Keldysh_times}
    |t-t'|=
    \begin{cases}
        |t-t'| & t,t'\text{ in forward branch} \\
        -|t-t'| & t,t' \text{ in backward branch} \\
        (t'-t) & 
        \begin{matrix} 
            \displaystyle t \text{ in forward branch, } \\
            \displaystyle t' \text{ in backward branch} 
        \end{matrix} \\
        (t-t') & 
        \begin{matrix}
            \displaystyle t \text{ in backward branch, } \\ 
            \displaystyle t' \text{ in forward branch.} 
        \end{matrix}
    \end{cases}
\end{equation}
Given these correlation functions, the current and noise are given by standard calculations of back-scattering currents (for example, see section 10 of Ref.~\cite{martin_noise_2005} and Ref.~\cite{chamon_tunneling_1995}).

\section{Exact solution}
\label{app:Exact}
\renewcommand{\theequation}{B\arabic{equation}}
\setcounter{equation}{0}

In this Appendix we follow Refs.~\cite{fendley_exact_PRL,fendley_exact_PRB,fendley_exact_sine_gordon,klassen_melzer_1990} to derive a TBA-based analysis of our model.

\subsection{Boundary Sine-Gordon}

Let us begin by observing the system near perfect normal reflection. From Eq.~\eqref{Unperturbed_Hamilonian} and Eq.~\eqref{QP_tunneling} we have the effective bosonic Hamiltonian
\begin{equation}
    \label{Exact_tunneling_Hamiltonian}
    \mathcal{H}=\frac{v m}{2\pi}\int_{-\infty}^0 dx \left[(\partial_x\theta)^2+(\partial_x\varphi)^2\right]+\tilde{B}\cos(\theta(0^-)),
\end{equation}
with the boundary condition $\varphi(x=0)=0$. Using the commutation relations between the bosonic fields, $[\theta(x),\varphi(x)]=i\frac{\pi}{m}\Theta(x'-x)$ we can perform a change of variables from this Hamiltonian to the standard boundary sine-Gordon form,

\begin{equation}
    \label{SG_Hamiltonian}
    \begin{matrix}
    \displaystyle \mathcal{H}=\frac{v}{2} \int_{-\infty}^0 dx \left[(\partial_x\phi)^2+\Pi^2+g\cos(\beta\phi)\right] \\ 
    \displaystyle +\tilde{B}\cos(\tfrac{1}{2}\beta\phi(0)).
    \end{matrix}
\end{equation}
In our Hamiltonian Eq.~\eqref{Exact_tunneling_Hamiltonian}, the bulk mass term $g$ goes to zero, and the other variables are given by
\begin{equation}
    \label{SG_variables_change}
    \begin{matrix}
    \displaystyle \phi\equiv\sqrt{\frac{m}{\pi}}\theta,
    \quad \Pi\equiv\sqrt{\frac{m}{\pi}}\partial_x\varphi, \quad \beta \equiv 2\sqrt{\frac{\pi}{m}}.
    \end{matrix}
\end{equation}
These fields are related to the total amount of charge carriers on the edges, defined in Eq.~\eqref{Total_charges}, via
\begin{equation}
    \label{Define_DN}
    \Delta N=N_L-N_R=\frac{1}{\pi}\int_{-\infty}^0 dx\partial_x\theta.
\end{equation}
The left-moving current given by Eq.~\eqref{Left_moving_current} becomes $\hat{I}= e \partial _t \Delta N - \frac{e^2V}{hm}$. 

We now describe the general solution of this Hamiltonian using the TBA equations. 

\subsection{Quasi-particle Scattering}

The spectrum of the boundary sine-Gordon Hamiltonian consists of a kink and an anti-kink, as well as $\lfloor{\lambda-1} \rfloor$ ``breather" states, where 
\begin{equation}
    \label{Define_lambda}
    \lambda \equiv \frac{8\pi}{\beta^2}-1=2m-1.
\end{equation}
We emphasize that $\lambda=2m-1$, whereas in Refs.~\cite{fendley_exact_PRL,fendley_exact_PRB}, $\lambda=m-1$. We therefore focus on the case of integer $\lambda$. The renormalized masses, $m_\pm$, of the kink $(+)$ and anti-kink $(-)$, and the renormalized masses of the breather states $m_b, (b=1,\dots,\lambda-1)$ are given by
\begin{equation}
    \label{breathers}
    m_\pm = \frac{M}{2},\quad m_b = M \sin\tfrac{b\pi}{2\lambda}.
\end{equation}
In the massless limit, $g \to 0$, the scale $M$ is arbitrary and cancels out of all physical results; for convenience, we henceforth take $M=2$. The massless excitations are hyper-relativistic quasi-particles, whose energies and momenta may hence be parametrized by their rapidities,
\begin{equation}
    E_j=p_j=m_j e^\theta, \quad j=1,\dots ,\lambda -1,+,-.
\end{equation}

From the sine-Gordon Hamiltonian Eq.~\eqref{SG_Hamiltonian}, we see that each kink/anti-kink corresponds to tunneling between minima of the massive-bulk term, $\phi \mapsto \phi \pm \frac{2\pi}{\beta}$. Expressed in terms of the original fields, this corresponds to $\theta \mapsto \theta \pm  \pi$, and a charge of $\Delta N= \pm 1$. Identifying kinks as unit positive charges and anti-kinks as unit negative charges allows us to obtain the chemical potentials of these excitations, $\mu_\pm=\pm eV$. The breathers carry no charge, and hence $\mu_b=0$ for all $b$.

We now treat our system as a scattering problem of these excitations off the boundary impurity at $x=0$. This scattering has been shown to happen one-by-one, with the probability for an incident kink to scatter either a kink or an anti-kink is given by the $S$ matrix elements $|S_{++}|^2+|S_{+-}|^2=1$~\cite{ghoshal_boundary_1994},
\begin{align}
    &|S_{++}(\theta-\theta_B)|^2=\frac{1}{1+e^{-2\lambda(\theta-\theta_B)}},\\
    &|S_{+-}(\theta-\theta_B)|^2=\frac{1}{1+e^{2\lambda(\theta-\theta_B)}}.
\end{align}

Here, $\theta_B$ is the back-scattering rapidity, obtained via the boundary energy scale in Eq.~\eqref{Crossover_scales} by $T_B=Te^{\theta_B}$, where $T$ is temperature. Each kink to anti-kink scattering event transfers a charge of $\Delta N=-2$, whereas kink to kink scattering events don't transfer charge.

\begin{figure}[t]
        \centering
        ~
        \newline
        \newline
         \includegraphics[width=0.4\textwidth]{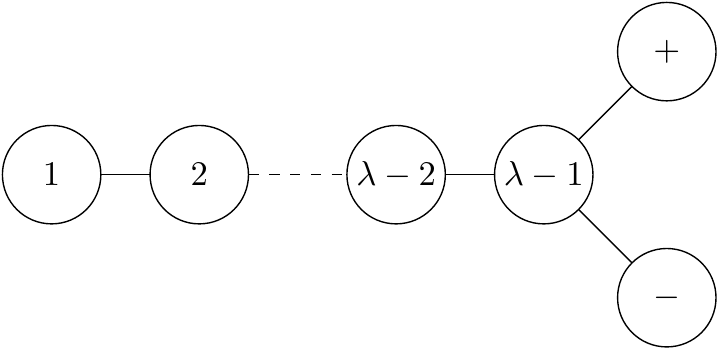}
         \caption{This diagram dictates which quasi-energies appear in which TBA equations via its incidence matrix $N$, \textit{cf.} Ref.~\cite{fendley_exact_PRB}. $N_{ij}=1$ for any two indices that are connected in the diagram, and $N_{ij}=0$ for any two indices that not connected. The incidence matrix for $\lambda=5$ is shown in Eq.~\eqref{Incidence_matrix}.}
         \label{fig:Incidence_Matrix}
\end{figure}

\subsection{Quasi-energies}

The one-by-one nature of the scattering enables us to define, for each type of excitation, an occupation number at each rapidity, $0 \leq f_j(\theta) \leq 1$. We parameterize this occupation number using a ``quasi-energy", $\epsilon_j(\theta)$, and the chemical potentials, $\mu_j$, such that
\begin{equation}
    f_j(\theta)=\frac{1}{1+e^{-\mu_j/T}e^{\epsilon_j(\theta)}}.
\end{equation}
This is analogous to the energy in Fermi-Dirac statistics, $\epsilon_j(\theta) \sim E_j(k)/T $. We proceed to define a density of states $n_j(\theta)$, 
\begin{equation}
    n_j(\theta)=\frac{T}{h} \frac{\partial \epsilon_j(\theta)}{\partial \theta}.
\end{equation}
Similarly, this is analogous to the density of states in the Fermi-Dirac case, $n_j(\theta)d\theta \sim \frac{\partial E_j(k)}{\partial k}\frac{dk}{h}$. This allows us to determine the total scattering current as
\begin{align}
    \label{Scattering_current_S_matrix}
    &\delta I=e\partial_t\Delta N=-2e\int_{-\infty}^{\infty}d\theta |S_{+-}(\theta-\theta_B)|^2\rho(\theta),\\
    &\rho(\theta) \equiv n_+(\theta)f_+(\theta)-n_-(\theta)f_-(\theta).
\end{align} 
Because we're near perfect normal reflection, we define $\delta I = I - \frac{e^2V}{hm}$. 

The quasi-energies satisfy a set of TBA equations~\cite{fendley_exact_PRB,fendley_exact_sine_gordon,ghoshal_boundary_1994,klassen_melzer_1990,zamolodchikov_thermodynamic_1991}
\begin{equation}
    \label{TBA_Equations}
    \epsilon_i(\theta)= \int_{-\infty}^{\infty}d\theta' K (\theta') \sum_j N_{ij}\ln(1+e^{-\mu_j/T}e^{\epsilon_j(\theta-\theta')}),
\end{equation}
with the boundary condition
\begin{equation}
    \epsilon_j(\theta\to\infty)\to m_j e^\theta.
\end{equation}
The kernel $K$ is given by
\begin{equation}
    K(\theta)=\frac{\lambda}{2\pi\cosh\lambda\theta},
\end{equation}
and $N_{ij}$ is the incidence matrix of the diagram presented in Fig.~\ref{fig:Incidence_Matrix}. Specifically, $N_{ij}=1$ for any two indices that are connected in the diagram, and $N_{ij}=0$ for any two indices that not connected. For example, for $m=3$, we have $\lambda = 5$, and the matrix $N$ is given by

\begin{equation}
    \label{Incidence_matrix}
    N = \bordermatrix{~ & 1 & 2 & 3 & 4 & + & - \cr
    1 & 0 & 1 & 0 & 0 & 0 & 0 \cr
    2 & 1 & 0 & 1 & 0 & 0 & 0 \cr
    3 & 0 & 1 & 0 & 1 & 0 & 0 \cr
    4 & 0 & 0 & 1 & 0 & 1 & 1 \cr
    + & 0 & 0 & 0 & 1 & 0 & 0 \cr
    - & 0 & 0 & 0 & 1 & 0 & 0 \cr
    }.
\end{equation}
The symmetries of the diagram in Fig.~\ref{fig:Incidence_Matrix} imply that $\epsilon_+(\theta)=\epsilon_-(\theta)\equiv\epsilon(\theta)$ and hence that $n_+(\theta)=n_-(\theta)\equiv n(\theta)$. The coupled integral equations Eq.~\eqref{TBA_Equations} asymptotically satisfy~\cite{klassen_melzer_1990}
\begin{align}
    \label{TBA_asymptotic}
    &\epsilon_j(\theta\to-\infty)\to\epsilon_j^0, \nonumber \\
    &\epsilon_\pm^0=\ln\lambda,\qquad\epsilon_b^0=\ln(b(b+2)),
\end{align} 
and may be solved iteratively using these asymptotic conditions. This gives full knowledge of the system. An efficient change of variables that prompts fast convergence is $z=e^\theta$, and $w=e^{-\theta'}$ such that
\begin{equation}
\epsilon_i(z)=\sum_j N_{ij}\frac{\lambda}{\pi}\int_{0}^{\infty}dw\frac{w^{\lambda-1}}{w^{2\lambda}+1}\ln(1+e^{-\mu_j/T}e^{\epsilon_j(z\cdot w)}).
\end{equation}

We are now in position to find an explicit solution to the current in terms of the quasi-energies.

\subsection{Physical observables}
Plugging all the above definitions into Eq.~\eqref{Scattering_current_S_matrix}, we finally find the main result of this Appendix,
\begin{widetext}
\begin{align}
    \label{Scattering_Current_Full}
    \delta I &= \frac{2eT}{h} \int_{-\infty}^{\infty}d\theta 
    \frac{1}{1+e^{2\lambda(\theta-\theta_B)}} \partial_\theta \ln \left(\frac{1+e^{eV/T}e^{-\epsilon(\theta)}}{1+e^{-eV/T}e^{-\epsilon(\theta)}} \right) \nonumber\\
    & = - \frac{2eT}{h} \ln \left(\frac{1+e^{eV/T}/\lambda}{1+e^{-eV/T}/\lambda} \right) + \frac{eT\lambda}{h}\int_{-\infty}^{\infty}d\theta \frac{1}{\cosh^2(\lambda(\theta-\ln(T_B/T)))} \ln \left(\frac{1+e^{eV/T}e^{-\epsilon(\theta)}}{1+e^{-eV/T}e^{-\epsilon(\theta)}} \right),
\end{align}
\end{widetext}
where to receive the second line we integrated by parts and utilized the asymptotic values Eq.~\eqref{TBA_asymptotic}.

We can immediately take the zero voltage limit and find the conductance $\delta G=\partial_V \delta I|_{V\to0}$ to be
\begin{align}
&\delta G (V\to0) =-\frac{4e^2}{h(\lambda+1)} +\frac{2e^2\lambda}{h} \times \nonumber\\
&\int_{-\infty}^{\infty} d\theta \frac{1}{\cosh^2(\lambda(\theta-\ln(T_B/T)))} \cdot \frac{1}{1+e^{\epsilon(\theta)|_{V\to0}}}
\end{align}
Interestingly, a complete solution for the quasi-energies can be found for $T\to0$. In this limit, only kinks proliferate the right-moving, positive-voltage edge, and the occupation factors become step functions. This leads to a hard cutoff, $A$, for the rapidities. The scattering current is now given by
\begin{equation}
    \delta I=-2e\int_{-\infty}^A d\theta|S_{+-}(\theta-\theta_B)|^2\rho(\theta),
\end{equation}
where now $f_-(\theta)=0$, and therefore $\rho(\theta) = n(\theta)f_+(\theta)$. 

Following the Weiner-Hopf technique \cite{fendley_exact_PRB,japaridze_exact_1984}, one can obtain an explicit expression for the scattering current,
\begin{equation}
    \label{Backscattering_rapidity_space}
    \delta I=-\frac{2e^2V}{h}\int_{-\infty}^0d\theta F(\theta)\frac{1}{1+e^{2\lambda(\theta+\ln(eV/\bar{T}_B))}},
\end{equation} 
where $F$ is given via
\begin{equation}
    F(\theta)=\int_{-\infty}^{\infty}\frac{d\alpha}{2\pi}e^{-i\alpha\theta}\tilde{F}(\alpha),\quad
    \tilde{F}(\alpha)=\frac{G_-(\alpha)G_+(0)}{1+i\alpha},
\end{equation}
and $\bar{T}_B$ is given via 
\begin{equation}
    \bar{T}_B=T_B\frac{G_+(i)}{G_+(0)}.
\end{equation}
The functions $G_\pm$ used to obtain these variables, derived via the Wiener-Hopf technique~\cite{japaridze_exact_1984}, are
\begin{equation}
        G_+(\alpha)=\sqrt{2\pi(\lambda+1)} \frac{\Gamma(-i\frac{(\lambda+1)\alpha}{2\lambda})} {\Gamma(-i\frac{\alpha}{2\lambda})\Gamma(\frac{1}{2}-i\frac{\alpha}{2})} e^{-i\alpha\Delta}
\end{equation}
\begin{equation}
         G_-(-\alpha)=G_+(\alpha),
\end{equation}
\begin{equation}
    \Delta=\frac{1}{2}\ln\lambda-\frac{\lambda+1}{2\lambda}\ln(\lambda+1).
\end{equation}
The hard cutoff, $A$, can also be described in terms of the functions $G_\pm$,
\begin{equation}
  A=\ln\frac{eV G_+(0)}{T G_+(i)}.  
\end{equation}

A numerically efficient way to evaluate the integral in Eq.~\eqref{Backscattering_rapidity_space} is given by 
\begin{equation}
 \delta I=-\frac{4e^2V}{h(\lambda+1)}+\frac{2e^2V}{h}\int_{-\infty}^{\infty}d\alpha\tilde{F}(\alpha)R(\alpha),   
\end{equation}
using the definition
\begin{equation}
    R(\alpha)\equiv\int_{-\infty}^{\infty}\frac{d\theta}{2\pi} e^{-i\alpha\theta}\frac{\Theta(-\theta)}{1+e^{-2\lambda(\theta+\ln(eV/\tilde{T}_B))}}.
\end{equation}
A similar expression can be derived and evaluated for the conductance.
\end{document}